\documentclass{elsart}

\usepackage{natbib}

 \usepackage{graphicx}
\setkeys{Gin}{clip=false,keepaspectratio=true,height=\textwidth,width=\textwidth}

\usepackage{xspace}

\begin{document}

%%% Macros

%% Useful astronomical symbols
\def\sq{\hbox{\rlap{$\sqcap$}$\sqcup$}}
\def\deg{\hbox{$^\circ$}}
\def\sqdeg{\sq\deg\xspace}
\def\arcmin{\hbox{$^\prime$}}
\def\arcsec{\hbox{$^{\prime\prime}$}}

\begin{frontmatter}

% Title, authors and addresses

% use the thanksref command within \title, \author or \address for footnotes;
% use the corauthref command within \author for corresponding author footnotes;
% use the ead command for the email address,
% and the form \ead[url] for the home page:
% \title{Title\thanksref{label1}}
% \thanks[label1]{}
% \author{Name\corauthref{cor1}\thanksref{label2}}
% \ead{email address}
% \ead[url]{home page}
% \thanks[label2]{}
% \corauth[cor1]{}
% \address{Address\thanksref{label3}}
% \thanks[label3]{}

\title{The Nearby Supernova Factory}

% use optional labels to link authors explicitly to addresses:
% \author[label1,label2]{}
% \address[label1]{}
% \address[label2]{}

\author{W. M. Wood-Vasey\corauthref{NSF}},
\corauth[NSF]{WMWV was supported in part by an NSF Graduate Research Fellowship.}
\ead{wmwood-vasey@lbl.gov}
\ead[url]{http://snfactory.lbl.gov}
\author{G. Aldering},
\author{B. C. Lee},
\author{S. Loken},
\author{P. Nugent},
\author{S. Perlmutter},
\author{J. Siegrist},
\author{L. Wang}
\address{Lawrence Berkeley National Laboratory, One Cyclotron Road, Mailstop 50R232, Berkeley, CA, 94720, USA}

\author{P. Antilogus},
\author{P. Astier},
\author{D. Hardin},
\author{R. Pain}
\address{Laboratoire de Physique Nucleaire et de Haute Energies de Paris, Paris, France (LPNHE)}

\author{Y. Copin},
\author{G. Smadja},
\author{E. Gangler},
\author{A. Castera}
\address{Institut de Physique Nucleaire de Lyon (IPNL), Lyon, France}

\author{G. Adam},
\author{R. Bacon},
\author{J-P. Lemonnier},
\author{A. Pecontal},
\author{E. Pecontal}
\address{Centre de Recherche Astronomique de Lyon (CRAL), Lyon, France}
%and
\author{R. Kessler}
\address{University of Chicago, Chicago, Illinois, USA}

\begin{abstract}
The Nearby Supernova Factory (SNfactory) is an ambitious project to
find and study in detail approximately 300 nearby Type Ia supernovae (SNe~Ia)
at redshifts \linebreak
\mbox{$0.03<z<0.08$}.  This program will provide an
exceptional data set of well-studied SNe in the nearby smooth
Hubble flow that can be used as calibration for the current and
future programs designed to use SNe to measure the cosmological
parameters.
The first key ingredient for this program is a reliable supply of
Hubble-flow SNe systematically discovered in unprecedented
numbers using the same techniques as those used in distant SNe
searches.  In 2002, 35 SNe were found using our test-bed
pipeline for automated SN search and discovery. The pipeline
uses images from the asteroid search conducted by the Near Earth
Asteroid Tracking group at JPL.  Improvements in our subtraction
techniques and analysis have allowed us to increase our effective
SN discovery rate to $\sim$12 SNe/month in 2003.

\end{abstract}


\end{frontmatter}

% main text
%%%%%%%%%%%%%%%%%%%%%%%%%%%%%%%%%%%%%%%%%

\section{Introduction}

Type Ia supernovae (SNe~Ia) have proven extremely useful as
standardizable candles to explore the expansion history of the
Universe~\citep{perlmutter97,perlmutter98a,garnavich98,riess98,perlmutter99}.
% RRR 
Ambitious follow-on experiments
are just starting, SNLS~\citep{pain03}, ESSENCE~\citep{garnavich02},
or have been planned, SNAP~\citep{aldering02a},
to extend the revolutionary result that the Universe is accelerating
to precise statements about the constituents and history of the
Universe.  However, a key assumption underlying these experiments is
that the current observed diversity in SNe~Ia will be well-behaved and
calibrated to allow for the desired precision measurements.
% RRR
This assumption can be tested in large part using nearby SNe~Ia.
Moreover, there is the possibility for such nearby studies to uncover
new relationships that will make SNe~Ia even better
standard candles, much as the width-luminosity relation has brought
SNe~Ia to their already impressive level of standardization.

The Nearby Supernova Factory (SNfactory) 
project is designed to bring this improved understanding of
SNe~Ia~\citep{aldering02b, pecontal03,lantz03}.  Over the course of
three years, it will study 300~SNe~Ia in the nearby smooth Hubble
flow.  These SNe will be observed with a dedicated instrument, the
SuperNova Integral Field Spectrograph (SNIFS), which is currently in
the final stages of construction.  SNIFS will provide simultaneous
spectrophotometric coverage of both the SN and the host galaxy at
3--5~day intervals during the rise and fall of each SN.  This
unprecedented dataset will provide a wealth of information on SNe Ia
and allow for an improved calibration of SNe~Ia for use in cosmology.

\section{The Supernova Search Dataset}

The SNfactory searches for SNe using wide-field images obtained in
collaboration with the Near Earth Asteroid Tracking (NEAT)
group~\citep{pravdo99} at the Jet Propulsion
Laboratory (JPL).  In their quest for asteroids, the NEAT
group observes hundreds of fields each night by taking three images
of each field, spaced fifteen to thirty minutes apart, and searching for
objects that move by more than a couple of arcseconds over this
period.  The SNfactory uses this temporal spacing to
eliminate asteroids and reduce cosmic-ray contamination.
The NEAT observing program covers the observable sky from $-40$ to
$+40$ degrees in declination every one to two weeks.

Since 2001, the Palomar 1.2-m. Oschin telescope has been used 
%as a part of 
by
the SNfactory-NEAT collaboration.  The NEAT group outfitted this
telescope with an automated system for control and observations and
added a 3-chip, 3~\sqdeg~field-of-view CCD camera (NEAT12GEN2) at the
spherical focal plane of this Schmidt reflector.  We spent 2001
designing, testing, and verifying the search pipeline with
the large data stream from this telescope.  Full search operations
began in the fall of 2002 and continued through April of 2003 when the
NEAT12GEN2 camera was replaced by the Yale QUEST group~\citep{quest}
with a 112-chip, 9~\sqdeg~field-of-view camera (QUESTII) capable of
both drift-scan and point-and-track observations.  QUESTII became
operational in August 2003 and is now providing data for the SNfactory search.  The NEAT group also uses the Haleakala
1.2-m MSSS telescope, but as this telescope has a smaller field of
view and poor image quality, the SNfactory has focused its search
efforts on the images from the Palomar Oschin telescope.

\section{Data Processing}

The SNfactory, in collaboration with the High Performance Wireless
Research and Education Network~\citep{hpwren}, has
established a high-speed, 6~Megabyte-per-second (MBps) radio internet
link to the San Diego Supercomputer Center (SDSC) from the Palomar
observatory.  Images are transmitted from the telescope and stored at
the National Energy Research Supercomputing Center (NERSC) High
Performance Storage System (HPSS).  The bandwidth from SDSC to
NERSC allows for near real-time transfer of 20--50~GB per night. 
%(there is a short delay needed for image compression).
%This transfer setup has been working continuously since August 2001.
 From HPSS the data are transferred to the 200-node NERSC Parallel
Distributed Systems Facility (PDSF) 
%for processing. 
and
% This cluster comprises approximately 200 dual 1-GHz PIII PCs.
%each with 2~GB of memory and 50~GB of scratch disk space.  
%The images, divided into groups based on the dark frame taken
%closest in time, 
are submitted for simultaneous processing on the PDSF cluster
in groups based on the dark frame taken
closest in time.
% Up to 15 processing jobs are run simultaneously.
%After processing, the fully reduced images 
The processed images are then registered with our
image database, renamed to match our canonical name format,
and saved to HPSS.
%  They are then moved back to central cluster storage so that
% they can be accessed by other processes.  After all of the images have
% been processed and moved to central cluster storage, they are saved to
% long-term storage on HPSS.

The SNfactory searches these data for SNe using 
%the technique of 
image subtraction.  
% Reference images of a given field from previous
% years of NEAT data are subtracted from the new image, and the resulting
% subtracted images are scanned for any remaining objects.  
The computers use a
sophisticated suite of image tools, but there
remains a significant, although ever-decreasing, amount of human interpretation needed to discriminate
the good SN candidates from the bad.
% Almost a decade of work has gone into our subtraction software, and it
% is continually being improved and rewritten.
The same software used by the SNfactory to search for nearby (z $\le
0.1$) SNe is used by the Supernova Cosmology Project (SCP) to search
for distant SNe.
%\subsection{Subtractions}
Our subtraction software begins by spatially registering all of the
images of a given field to a common reference system.
The images to be used as a reference are
%shifted to line up with the designated primary reference images
%and then 
%added together,
stacked into a single coadded image,
while the images to be searched are coadded into
%The list of images to be searched is split into two parts,
%{\em new1} and {\em new2},
two separate images to allow for later checks 
for asteroids and cosmic rays. 
To account for differences in the effective point-spread-function from
variations in atmospheric and telescope conditions, we
calculate a convolution kernel to match each coadded image
to the worst-seeing coadded image of the set. 
%% RRR
%Once the images have been matched through the convolution kernel, the
%coadded reference image is subtracted from the two search images,
%{\em new1} and {\em new2}.
%to generate separate {\em sub1} and {\em sub2} scores for each
%candidate.  
%The final subtracted image is then simply {\em sub}$=${\em new1}$+${\em new2}$-${\em ref}.
%The {\em sub} image is derived from the sum of {\em new1}
%and {\em new2} into {\em new}, which is separately convolution-matched
%with the other images and then subtracted by the convolved reference
%to yield the final subtracted image.
%\subsection{Automated Scanning}
%% RRR
%Once the images are convolution-matched and subtracted, a
An automated scanning program
% takes the full subtractions and 
looks for objects in the convolution-matched, subtracted
image and applies a variety of selection criteria 
to eliminate cosmic rays, asteroids, and subtraction and detector artifacts.
This program compiles a list of interesting candidates to be looked at in more
detail by a human scanner.
% All processing up to this point is completely automated
% and runs daily with minimal human supervision.
Every day, human scanners consult the list of interesting candidates
and decide whether or not a computer-flagged candidate appears to be a real, variable object.
%\subsection{Cross Checks}
% To check for known cases of variable objects that are not SNe,
% we consult the Minor Planet Center catalog~\citep{mpchecker} and the APS
% catalog~\citep{cabanela03}.  In addition, we check past NEAT images
% of the candidate to verify that its lightcurve is
% consistent with that of a SN.  This last step also allows us to
% constrain the discovery epoch for a new SN.
%\subsection{Confirmation Image}
Once we have a promising candidate, we submit it to the
target list for the next night of observation with NEAT.  
% Our current turn-around time is 2~days; we will decrease that to
% just 1~day by the end of 2004.
% Once those data come in, 
% we prioritize the images and subtractions to look at that same region
% of sky, verifying that the variable object is still there.
%\subsection{Report Discovery}
After obtaining confirmation images of a candidate that
reveal it to exhibit the appropriate behavior for a SN, we announce
the apparent SN in the IAU Circulars.  A confirmation spectrum is
desirable but currently not always possible.  When the SNIFS
instrument is installed on the Hawaii 2.2-m telescope, it will be
automatically scheduled to confirm and follow SNe.  Spectra will be
taken of each SN at 3--5~day intervals over a span of roughly 60~days.

\section{Results}

%\subsection{Supernovae found to date}

\begin{figure}
\includegraphics[width=6in,height=4in]{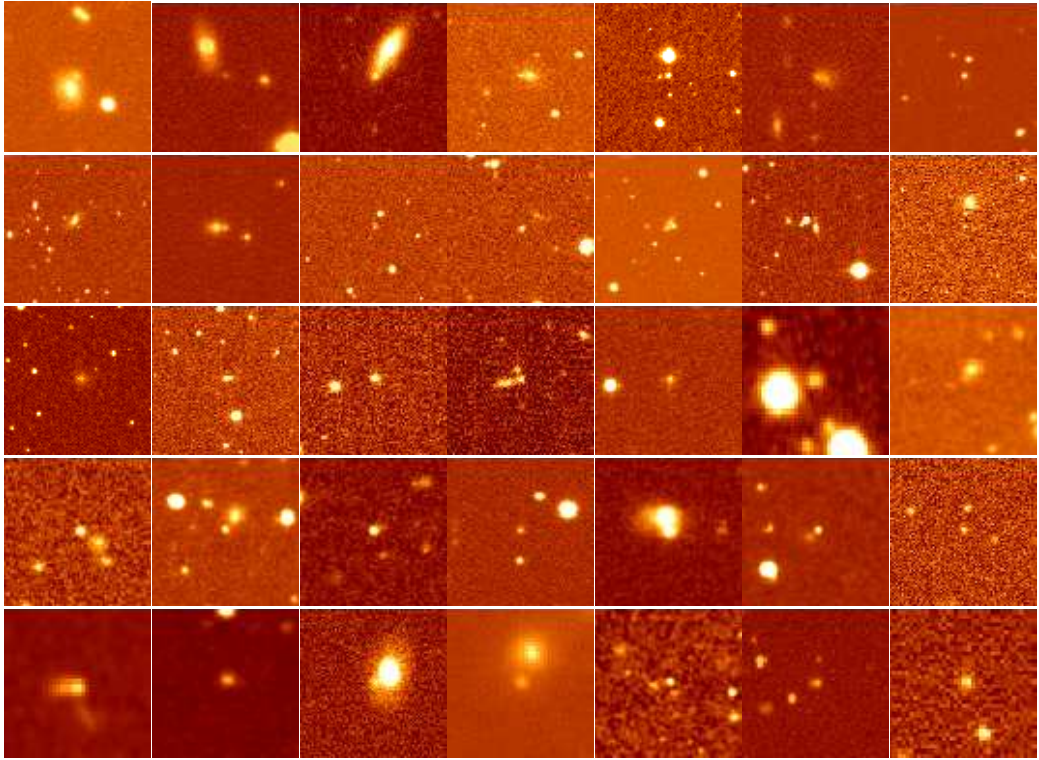}
\caption{
A mosaic of the 35 supernovae found by the Nearby Supernova
Factory pipeline in 2002.  Each image is centered on the respective supernova.
}
\label{fig:sne_mosaic}
\end{figure}

Eighty-three SNe have been found using the techniques described
above and have been accepted by the International Astronomical Union.
Fig.~\ref{fig:sne_mosaic} shows the 35 SNe we discovered in
2002.  In the first five months of 2003, we found and reported an
additional 48 SNe.  In addition, our search identified
another 17 SNe that had already been reported by other groups.
We are currently running with a very conservative set of selection
criteria and need human eyes to scan $\sim5$\% of the successful
subtractions for each night.  However, many of the SNe discovered in
2003 could have been found using very restrictive criteria that
would have required human scanning for $<1$\% of the subtractions.  
To further reduce our scanning
burden, efforts are ongoing to understand the parameter space of our
candidate scores to uniquely identify the SNe.
% and distinguish them from
% asteroids, variable stars, and image and subtraction artifacts.

\begin{figure}
\resizebox{12cm}{!}{
\includegraphics[angle=270]{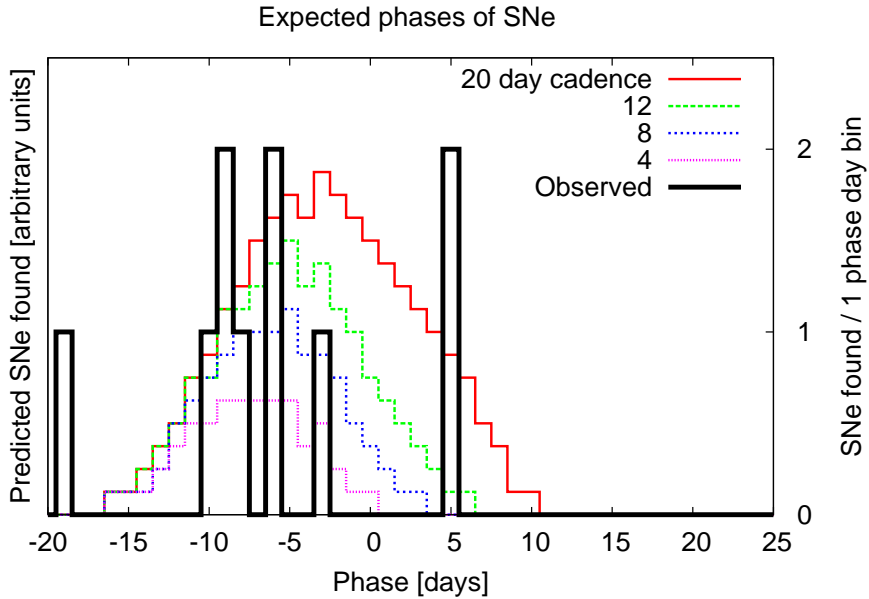}
}
\caption{
The distribution of discovery epoch for the SNe~Ia with determined dates of maximum found in the
SNfactory dataset compared with the
simulations described in the text.  The model curves are not calculated in
absolute units and have been scaled for comparison with the
observations.  As the model curves show, a shorter cadence gives fewer
supernovae but better constraints on the epoch at discovery.  }
\label{fig:compare_phase}
\end{figure}

In order to understand our search results, we have developed a simple search
simulator.  We use a V-band lightcurve template from
\citet{goldhaber01} to model the rise and fall of SNe~Ia.\footnote{We
have found that the NEAT unfiltered magnitudes track V-band SN
lightcurves quite well.}  The simulator calculates the amount of time
a supernova would be visible at a given redshift assuming a normal
(stretch~$=1$) SN~Ia. We considered a redshift range from $0<z<0.2$,
a limiting unfiltered magnitude of 19.5, and 
several different repeat coverage cadences for our simulations.  We
included the typical NEAT sky coverage rate, $S_D = 500~\sqdeg$/night,
and assumed $S_O=$10,000$~\sqdeg$ of usable sky a night.  There is a
maximum effective cadence, $C_\mathrm{max}$, beyond which one is just
idling the telescope: $C_\mathrm{max} = S_O / ( S_D -
\frac{S_0}{365~\mathrm{days}})$.  
% For a sky coverage of 500~\sqdeg a
% night and $\sim$10,000~\sqdeg of usable sky per night, 
For the NEAT observation program, this 
cadence is approximately 20 days.  
% Thus, this is the maximum cadence shown in Fig.~\ref{fig:compare_phase}.
%\subsection{Discovery Epoch}
An important goal of the SNfactory is to discover SNe early enough
after explosion to follow them through their rise and fall.
Fig.~\ref{fig:compare_phase} shows a comparison of our simulations of
the SNe discovery phase with
the actual discovery phase for SNe from the SNfactory search having
well-known dates of maximum.  For shorter sky coverage cadences
less sky can be covered and so fewer SNe are discovered overall,
although the epoch of each SN can be better constrained.
Fig.~\ref{fig:compare_phase} clearly shows that the SNfactory search
pipeline is successfully finding SNe~Ia early in their lightcurves.

% The number of SNe found at different redshifts (see
% Fig.~\ref{fig:sweet_spot}) is a product of the increasing volume with
% redshift and the decreasing time a SN is observable during its rise
% and fall.  The observable time is also modulated by the SN lightcurve
% width-brightness correlation.  At lower redshifts the increasing
% volume dominates and the number of SNe discovered at a given redshift
% increases with redshift.  But, as the redshift nears the limiting
% redshift for a SN to be observable at peak, the number of SNe
% discovered smoothly rolls over and falls to zero. Similarly, the epoch
% at which a SN is discovered is constrained by its redshift: at early
% times a SN is visible out to a smaller redshift than at maximum.  This
% effect is modeled in Fig.~\ref{fig:compare_phase} to show the
% predicted phase of discovery for SNe~Ia.

\section{Conclusion}

The Nearby Supernova Factory search pipeline is operational and has
proven the ability to discover $\sim12$~SNe/month $\Rightarrow$
$\sim$150~SNe/year.  As $\frac{2}{3}$ of the supernovae discovered in
our search have been SNe~Ia, we expect to discover $\sim100$~SNe~Ia/year.
Most of these supernovae have been discovered sufficiently early to enable
detailed study starting before maximum light.  This extensive study will enable
improvements in the use of SNe~Ia for cosmological measurements,
and provide a wealth of information on the supernovae themselves.

%% Figures

\begin{figure}
\resizebox{12cm}{!}{
\includegraphics[angle=270]{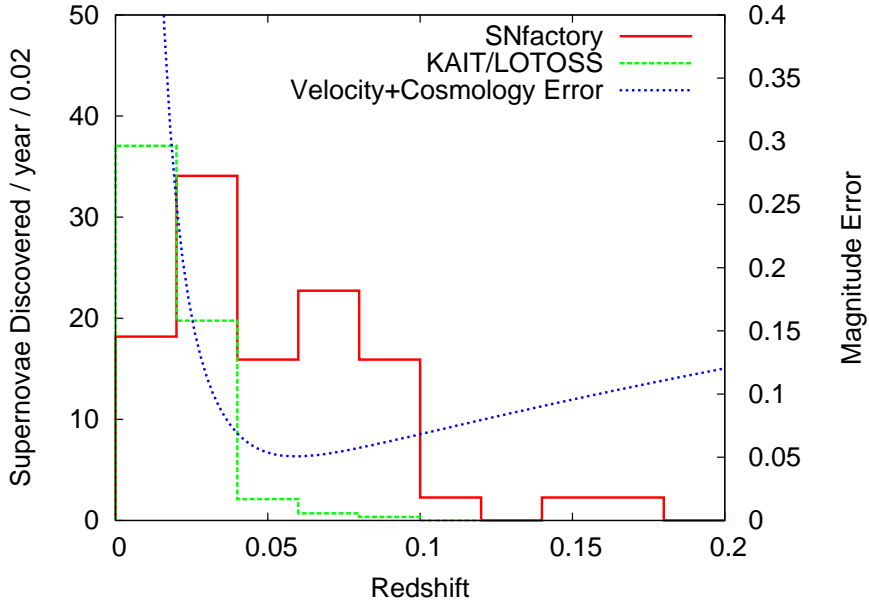}
}
\caption{
The SNfactory is operating in the ``sweet spot'' redshift range
between peculiar-velocity noise and cosmological uncertainty.  The
SNfactory curve is the redshift distribution of supernovae found and
spectroscopically confirmed in our search to date scaled up to 100
SNe/year.  The velocity error is for an assumed $300$~km/s velocity
dispersion.  The cosmology error is modeled as the difference between
an Einstein-de Sitter cosmology and a Universe with $\Omega_M=0.3$ and
$\Omega_\Lambda=0.7$.}
\label{fig:sweet_spot}
\end{figure}

%\bibliographystyle{astroads}
%%\bibliographystyle{apj}
%\bibliography{apj-jour,thesis_refs}

\begin{thebibliography}{16}
\expandafter\ifx\csname natexlab\endcsname\relax\def\natexlab#1{#1}\fi
\expandafter\ifx\csname href\endcsname\relax
  \def\href#1#2{}\fi
\expandafter\ifx\csname urllinklabel\endcsname\relax
  \def\urllinklabel{[LINK]}\fi
\expandafter\ifx\csname adsurllinklabel\endcsname\relax
  \def\adsurllinklabel{[ADS]}\fi

\bibitem[{{Aldering} {et~al.}(2002{\natexlab{a}}){Aldering}, {Adam},
  {Antilogus}, {Astier}, {Bacon}, {Bongard}, {Bonnaud}, {Copin}, {Hardin},
  {Henault}, {Howell}, {Lemonnier}, {Levy}, {Loken}, {Nugent}, {Pain},
  {Pecontal}, {Pecontal}, {Perlmutter}, {Quimby}, {Schahmaneche}, {Smadja}, \&
  {Wood-Vasey}}]{aldering02a}
{Aldering}, G., {Adam}, G., {Antilogus}, P., {Astier}, P., {Bacon}, R.,
  {Bongard}, S., {Bonnaud}, C., {Copin}, Y., {Hardin}, D., {Henault}, F.,
  {Howell}, D.~A., {Lemonnier}, J., {Levy}, J., {Loken}, S.~C., {Nugent},
  P.~E., {Pain}, R., {Pecontal}, A., {Pecontal}, E., {Perlmutter}, S.,
  {Quimby}, R.~M., {Schahmaneche}, K., {Smadja}, G., \& {Wood-Vasey}, W.~M.
  2002{\natexlab{a}}, in Survey and Other Telescope Technologies and
  Discoveries. Edited by Tyson, J. Anthony; Wolff, Sidney. Proceedings of the
  SPIE, Volume 4836, pp. 61-72 (2002)., 61--72


\bibitem[{{Aldering} {et~al.}(2002{\natexlab{b}}){Aldering}, {Akerlof},
  {Amanullah}, {Astier}, {Barrelet}, {Bebek}, {Bergstrom}, {Bercovitz},
  {Bernstein}, {Bester}, {Bonissent}, {Bower}, {Carithers}, {Commins}, {Day},
  {Deustua}, {DiGennaro}, {Ealet}, {Ellis}, {Eriksson}, {Fruchter}, {Genat},
  {Goldhaber}, {Goobar}, {Groom}, {Harris}, {Harvey}, {Heetderks}, {Holland},
  {Huterer}, {Karcher}, {Kim}, {Kolbe}, {Krieger}, {Lafever}, {Lamoreux},
  {Lampton}, {Levi}, {Levin}, {Linder}, {Loken}, {Malina}, {Massey}, {McKay},
  {McKee}, {Miquel}, {Moertsell}, {Mostek}, {Mufson}, {Musser}, {Nugent},
  {Oluseyi}, {Pain}, {Palaio}, {Pankow}, {Perlmutter}, {Pratt}, {Prieto},
  {Refregier}, {Rhodes}, {Robinson}, {Roe}, {Sholl}, {Schubnell}, {Smadja},
  {Smoot}, {Spadafora}, {Tarle}, {Tomasch}, {von der Lippe}, {Vincent},
  {Walder}, \& {Wang}}]{aldering02b}
{Aldering}, G., {Akerlof}, C.~W., {Amanullah}, R., {Astier}, P., {Barrelet},
  E., {Bebek}, C., {Bergstrom}, L., {Bercovitz}, J., {Bernstein}, G.~M.,
  {Bester}, M., {Bonissent}, A., {Bower}, C., {Carithers}, W.~C., {Commins},
  E.~D., {Day}, C., {Deustua}, S.~E., {DiGennaro}, R.~S., {Ealet}, A., {Ellis},
  R.~S., {Eriksson}, M., {Fruchter}, A., {Genat}, J., {Goldhaber}, G.,
  {Goobar}, A., {Groom}, D.~E., {Harris}, S.~E., {Harvey}, P.~R., {Heetderks},
  H.~D., {Holland}, S.~E., {Huterer}, D., {Karcher}, A., {Kim}, A.~G., {Kolbe},
  W.~F., {Krieger}, B., {Lafever}, R., {Lamoreux}, J.~C., {Lampton}, M.~L.,
  {Levi}, M.~E., {Levin}, D.~S., {Linder}, E.~V., {Loken}, S.~C., {Malina}, R.,
  {Massey}, R., {McKay}, T., {McKee}, S.~P., {Miquel}, R., {Moertsell}, E.,
  {Mostek}, N., {Mufson}, S., {Musser}, J.~A., {Nugent}, P.~E., {Oluseyi},
  H.~M., {Pain}, R., {Palaio}, N.~P., {Pankow}, D.~H., {Perlmutter}, S.,
  {Pratt}, R., {Prieto}, E., {Refregier}, A., {Rhodes}, J., {Robinson}, K.~E.,
  {Roe}, N., {Sholl}, M., {Schubnell}, M.~S., {Smadja}, G., {Smoot}, G.~F.,
  {Spadafora}, A., {Tarle}, G., {Tomasch}, A.~D., {von der Lippe}, H.,
  {Vincent}, D., {Walder}, J.-P., \& {Wang}, G. 2002{\natexlab{b}}, in Future
  Research Direction and Visions for Astronomy. Edited by Dressler, Alan M.
  Proceedings of the SPIE, Volume 4835, pp. 146-157 (2002)., 146--157


\bibitem[{Braun(2003)}]{hpwren}
Braun, H.-W. 2003, {High Performance Wireless Research and Education Network}
 \href{http://hpwren.ucsd.edu/}{\urllinklabel}

\bibitem[{{Cabanela} {et~al.}(2003){Cabanela}, {Humphreys}, {Aldering},
  {Larsen}, {Odewahn}, {Thurmes}, \& {Cornuelle}}]{cabanela03}
{Cabanela}, J.~E., {Humphreys}, R.~M., {Aldering}, G., {Larsen}, J.~A.,
  {Odewahn}, S.~C., {Thurmes}, P.~M., \& {Cornuelle}, C.~S. 2003, \pasp, 115,
  837


\bibitem[{{Garnavich} {et~al.}(2002){Garnavich}, {Holland}, {Schmidt},
  {Krisciunas}, {Smith}, {Suntzeff}, {Becker}, {Miceli}, {Miknaitis}, {Rest},
  {Stubbs}, {Filippenko}, {Jha}, {Li}, {Challis}, {Kirshner}, {Matheson},
  {Barris}, {Tonry}, {Riess}, {Leibundgut}, {Sollerman}, {Spyromilio},
  {Clocchiatti}, {Pompea}, \& {High-Z Supernova Search Team}}]{garnavich02}
{Garnavich}, P.~M., {Holland}, S.~T., {Schmidt}, B.~P., {Krisciunas}, K.,
  {Smith}, R.~C., {Suntzeff}, N.~B., {Becker}, A., {Miceli}, A., {Miknaitis},
  G., {Rest}, A., {Stubbs}, C., {Filippenko}, A.~V., {Jha}, S., {Li}, W.,
  {Challis}, P., {Kirshner}, R.~P., {Matheson}, T., {Barris}, B., {Tonry},
  J.~L., {Riess}, A.~G., {Leibundgut}, B., {Sollerman}, J., {Spyromilio}, J.,
  {Clocchiatti}, A., {Pompea}, S., \& {High-Z Supernova Search Team}. 2002,
  Bulletin of the American Astronomical Society, 34, 1233


\bibitem[{{Garnavich} {et~al.}(1998){Garnavich}, {Jha}, {Challis},
  {Clocchiatti}, {Diercks}, {Filippenko}, {Gilliland}, {Hogan}, {Kirshner},
  {Leibundgut}, {Phillips}, {Reiss}, {Riess}, {Schmidt}, {Schommer}, {Smith},
  {Spyromilio}, {Stubbs}, {Suntzeff}, {Tonry}, \& {Carroll}}]{garnavich98}
{Garnavich}, P.~M., {Jha}, S., {Challis}, P., {Clocchiatti}, A., {Diercks}, A.,
  {Filippenko}, A.~V., {Gilliland}, R.~L., {Hogan}, C.~J., {Kirshner}, R.~P.,
  {Leibundgut}, B., {Phillips}, M.~M., {Reiss}, D., {Riess}, A.~G., {Schmidt},
  B.~P., {Schommer}, R.~A., {Smith}, R.~C., {Spyromilio}, J., {Stubbs}, C.,
  {Suntzeff}, N.~B., {Tonry}, J., \& {Carroll}, S.~M. 1998, \apj, 509, 74


\bibitem[{{Goldhaber} {et~al.}(2001){Goldhaber}, {Groom}, {Kim}, {Aldering},
  {Astier}, {Conley}, {Deustua}, {Ellis}, {Fabbro}, {Fruchter}, {Goobar},
  {Hook}, {Irwin}, {Kim}, {Knop}, {Lidman}, {McMahon}, {Nugent}, {Pain},
  {Panagia}, {Pennypacker}, {Perlmutter}, {Ruiz-Lapuente}, {Schaefer},
  {Walton}, \& {York}}]{goldhaber01}
{Goldhaber}, G., {Groom}, D.~E., {Kim}, A., {Aldering}, G., {Astier}, P.,
  {Conley}, A., {Deustua}, S.~E., {Ellis}, R., {Fabbro}, S., {Fruchter}, A.~S.,
  {Goobar}, A., {Hook}, I., {Irwin}, M., {Kim}, M., {Knop}, R.~A., {Lidman},
  C., {McMahon}, R., {Nugent}, P.~E., {Pain}, R., {Panagia}, N., {Pennypacker},
  C.~R., {Perlmutter}, S., {Ruiz-Lapuente}, P., {Schaefer}, B., {Walton},
  N.~A., \& {York}, T. 2001, \apj, 558, 359
 \href{http://adsabs.harvard.edu/cgi-bin/nph-bib_query?bibcode=2001ApJ...558..%
359G&db_key=AST}{\adsurllinklabel}

\bibitem[{{International Astronomical Union}(2003)}]{mpchecker}
{International Astronomical Union}. 2003, {MPChecker: Minor Planet Checker}
 \href{http://scully.harvard.edu/~cgi/CheckMP}{\urllinklabel}

\bibitem[{{Lantz}(2003)}]{lantz03}
{Lantz}, B.~{\em et~al.}. 2003, in {Proceedings of the SPIE}


\bibitem[{{P{\' e}contal} {et~al.}(2003){P{\' e}contal}, {Aldering}, {Adam},
  {Antilogus}, {Astier}, {Copin}, {H{\' e}nault}, {Lemonnier}, {Nugent},
  {Pain}, {P{\' e}contal}, {Perlmutter}, {Quimby}, {Smadja}, \&
  {Wood-Vasey}}]{pecontal03}
{P{\' e}contal}, E., {Aldering}, G., {Adam}, G., {Antilogus}, P., {Astier}, P.,
  {Copin}, Y., {H{\' e}nault}, F., {Lemonnier}, J.-P., {Nugent}, P., {Pain},
  R., {P{\' e}contal}, A., {Perlmutter}, S., {Quimby}, R., {Smadja}, G., \&
  {Wood-Vasey}, M. 2003, in From Twilight to Highlight: The Physics of
  Supernovae. Proceedings of the ESO/MPA/MPE Workshop held in Garching,
  Germany, 29-31 July 2002, p. 404., 404--+


\bibitem[{{Pain} \& {The Snls Collaboration}(2003)}]{pain03}
{Pain}, R. \& {The Snls Collaboration}. 2003, in From Twilight to Highlight:
  The Physics of Supernovae. Proceedings of the ESO/MPA/MPE Workshop held in
  Garching, Germany, 29-31 July 2002, p. 408., 408--+


\bibitem[{{Perlmutter}(1998)}]{perlmutter98b}
{Perlmutter}, S. 1998, Bulletin of the American Astronomical Society, 30, 1388


\bibitem[{{Perlmutter} {et~al.}(1999){Perlmutter}, {Aldering}, {Goldhaber},
  {Knop}, {Nugent}, {Castro}, {Deustua}, {Fabbro}, {Goobar}, {Groom}, {Hook},
  {Kim}, {Kim}, {Lee}, {Nunes}, {Pain}, {Pennypacker}, {Quimby}, {Lidman},
  {Ellis}, {Irwin}, {McMahon}, {Ruiz-Lapuente}, {Walton}, {Schaefer}, {Boyle},
  {Filippenko}, {Matheson}, {Fruchter}, {Panagia}, {Newberg}, {Couch}, \& {The
  Supernova Cosmology Project}}]{perlmutter99}
{Perlmutter}, S., {Aldering}, G., {Goldhaber}, G., {Knop}, R.~A., {Nugent}, P.,
  {Castro}, P.~G., {Deustua}, S., {Fabbro}, S., {Goobar}, A., {Groom}, D.~E.,
  {Hook}, I.~M., {Kim}, A.~G., {Kim}, M.~Y., {Lee}, J.~C., {Nunes}, N.~J.,
  {Pain}, R., {Pennypacker}, C.~R., {Quimby}, R., {Lidman}, C., {Ellis}, R.~S.,
  {Irwin}, M., {McMahon}, R.~G., {Ruiz-Lapuente}, P., {Walton}, N., {Schaefer},
  B., {Boyle}, B.~J., {Filippenko}, A.~V., {Matheson}, T., {Fruchter}, A.~S.,
  {Panagia}, N., {Newberg}, H.~J.~M., {Couch}, W.~J., \& {The Supernova
  Cosmology Project}. 1999, \apj, 517, 565
 \href{http://adsabs.harvard.edu/cgi-bin/nph-bib_query?bibcode=1999ApJ...517..%
565P&db_key=AST}{\adsurllinklabel}

\bibitem[{{Perlmutter} {et~al.}(1997){Perlmutter}, {Gabi}, {Goldhaber},
  {Goobar}, {Groom}, {Hook}, {Kim}, {Kim}, {Lee}, {Pain}, {Pennypacker},
  {Small}, {Ellis}, {McMahon}, {Boyle}, {Bunclark}, {Carter}, {Irwin},
  {Glazebrook}, {Newberg}, {Filippenko}, {Matheson}, {Dopita}, {Couch}, \& {The
  Supernova Cosmology Project}}]{perlmutter97}
{Perlmutter}, S., {Gabi}, S., {Goldhaber}, G., {Goobar}, A., {Groom}, D.~E.,
  {Hook}, I.~M., {Kim}, A.~G., {Kim}, M.~Y., {Lee}, J.~C., {Pain}, R.,
  {Pennypacker}, C.~R., {Small}, I.~A., {Ellis}, R.~S., {McMahon}, R.~G.,
  {Boyle}, B.~J., {Bunclark}, P.~S., {Carter}, D., {Irwin}, M.~J.,
  {Glazebrook}, K., {Newberg}, H.~J.~M., {Filippenko}, A.~V., {Matheson}, T.,
  {Dopita}, M., {Couch}, W.~J., \& {The Supernova Cosmology Project}. 1997,
  \apj, 483, 565
 \href{http://adsabs.harvard.edu/cgi-bin/nph-bib_query?bibcode=1997ApJ...483..%
565P&db_key=AST}{\adsurllinklabel}

\bibitem[{{Riess} {et~al.}(1998){Riess}, {Filippenko}, {Challis},
  {Clocchiatti}, {Diercks}, {Garnavich}, {Gilliland}, {Hogan}, {Jha},
  {Kirshner}, {Leibundgut}, {Phillips}, {Reiss}, {Schmidt}, {Schommer},
  {Smith}, {Spyromilio}, {Stubbs}, {Suntzeff}, \& {Tonry}}]{riess98}
{Riess}, A.~G., {Filippenko}, A.~V., {Challis}, P., {Clocchiatti}, A.,
  {Diercks}, A., {Garnavich}, P.~M., {Gilliland}, R.~L., {Hogan}, C.~J., {Jha},
  S., {Kirshner}, R.~P., {Leibundgut}, B., {Phillips}, M.~M., {Reiss}, D.,
  {Schmidt}, B.~P., {Schommer}, R.~A., {Smith}, R.~C., {Spyromilio}, J.,
  {Stubbs}, C., {Suntzeff}, N.~B., \& {Tonry}, J. 1998, \aj, 116, 1009


\bibitem[{{The QUEST Collaboration}(2003)}]{quest}
{The QUEST Collaboration}. 2003, The Palomar-QUEST Variability Survey
 \href{http://hepwww.physics.yale.edu/quest/palomar.html}{\urllinklabel}

\end{thebibliography}
%\begin{thebibliography}{}

% \bibitem[]{} [Names(Year)]{label} or \bibitem[]{} [Names(Year)Long names]{label}.
% (\harvarditem{Name}{Year}{label} is also supported.)
% Text of bibliographic item

%\bibitem[Alcock et al (2001)]{Alcock:2001} Alcock C. et al: 2001, ApJS 136, 439
%\bibitem[Bahcall (1984)]{Bahcall:1984} Bahcall J.N.: 1984, ApJ 287, 926
%\bibitem[Schombert et al (2001)]{Schombert:2001} Schombert, J.M., McGaugh, S. S., Eder, J. A.: 2001, AJ 121, 2420 

%\end{thebibliography}

\end{document}